\documentclass[%
 reprint,
 superscriptaddress,
 onecolumn,
nobibnotes,
 amsmath,amssymb,
 aps,
 pre,
]{revtex4-2}

\usepackage{graphicx}
\usepackage{dcolumn}
\usepackage{bm}

\begin{document}

\preprint{APS/123-QED}

\title{Spin Transport Hydrodynamics of Polarized Deuterium-Tritium Fusion Plasma}

\author{Ronghao Hu}
\author{Hao Zhou}
\author{Zhihao Tao}
\author{Zhihao Zhang}
\author{Meng Lv}
\email{Corresponding author, lvmengphys@scu.edu.cn}
\affiliation{College of Physics, Sichuan University, Chengdu, 610065, People’s Republic of China}
\affiliation{Key Laboratory of High Energy Density Physics and Technology, Ministry of Education, Sichuan University, Chengdu, 610064, People’s Republic of China}

\author{Shiyang Zou}
\author{Yongkun Ding}
\affiliation{Institute of Applied Physics and Computational Mathematics, Beijing, 100094, People's Republic of China}
%


\begin{abstract}
The spin transport equations for polarized deuterium-tritium (DT) fusion plasma are derived with the density matrix formulation, which are used to investigate the hydrodynamics of polarized DT-gas-filled targets during indirectly driven inertial confinement fusion implosions. The depolarization of DT ions by strong self-generated magnetic fields can be captured by the spin transport equation. The neutron yield, angular distribution and neutron beam polarization are obtained from three-dimensional spin transport hydrodynamics simulations of the target implosions. The simulation results indicate that an optimized spin alignment of the polarized target can reduce the depolarization of DT ions and the neutron beams induced by polar mode asymmetries in indirectly driven implosions.
\end{abstract}

\maketitle

\section{Introduction}
Inertial confinement fusion (ICF) is a promising approach to produce controlled burning plasmas and high flux neutron beams in laboratory \cite{betti2016,*hurricane2014,*lepape2018,*zylstra2021,*zylstra2022}. Using spin-polarized fuels in ICF can potentially enhance the neutron yield and modify the angular distribution of the neutron beam, and more importantly generate spin-polarized neutron beam in certain emission direction \cite{kulsrud1982,*kulsrud1986,more1983,temporal2012,ciullo2016}. Polarized neutron scattering and polarized neutron imaging are indispensable tools to probe the structure and dynamics of magnetic systems \cite{kardjilov2008,*lazpita2017,*hilger2018}. Increasing the flux of the polarized neutron source can shorten the time required to obtain high quality scattering signals and neutron images. Both deuterium-tritium (DT) and deuterium-deuterium (DD) reactions can be used in ICF to generate polarized neutron beams \cite{perkins1961,*simmons1971}. However, the key nuclear physics data, such as fusion cross-section, neutron angular distribution and neutron polarization for polarized DT and DD reactions at ICF relative conditions, lack experimental measurements \cite{ciullo2016}. For DT reaction, Kulsrud \emph{et al.} developed the theoretic framework to predict fusion cross-section, neutron angular distribution and neutron polarization for arbitrary DT polarizations \cite{kulsrud1982,kulsrud1986}. \emph{Ab initio} calculations can also be used to predict the polarized DT fusion cross-section and neutron angular distribution \cite{hupin2019}. For DD reaction, several models give inconsistent predictions of fusion cross-sections \cite{ciullo2016}. These nuclear physics data can be measured using ICF implosions \cite{zylstra2016,casey2017} as long as polarized targets can be assembled. The atomic beam source can generate polarized deuterium and tritium atoms with high polarization \cite{roberts1992,baumgarten2003}. The nuclear polarization of atoms can be preserved during recombination to form ``hyperpolarized molecules'' \cite{engels2015}. If the polarized gas can be filled into the ICF capsule without severe depolarization, then the most significant question remaining is whether the polarized fuel could survive in the ICF implosion and produce polarized neutron beam.

The major depolarization mechanism of polarized DT fuel during ICF implosion is magnetic field induced depolarization \cite{hu2020}. Hydrodynamic instabilities, like Rayleigh-Taylor instability (RTI) and Richtmyer-Meshkov instability, can generate intense magnetic fields due to the Biermann battery effect \cite{manuel2012,srinivasan2012,walsh2017}. The periods of the Larmor precession for DT nuclei in strong magnetic fields are close to the ICF confinement time, so the depolarization can not be neglected. Due to the smaller gyromagnetic ratio, deuterons can sustain a higher polarization than tritons during the implosion. The depolarization and spin transport process can be simulated using particle-based methods \cite{gong2020,*gong2021}, hydrodynamic methods \cite{marklund2007} or hybrid methods \cite{hu2020}. Hydrodynamics simulations are widely used to interpret ICF experiments \cite{clark2016}, but conventional hydrodynamics codes do not include spin transport simulation. For spin polarized fusion, the probability distributions for spin eigenstates of DT are necessary to obtain the fusion cross-section and neutron angular distrubtion. The previously proposed spin transport equation using the vector polarization is not enough for spin-1 particles, whose tensor polarization is also needed to obtain the probability distribution for spin eigenstates \cite{marklund2007,zhang1988,*balescu1988}. In this Letter, we present the unified spin transport equation for spin-$\frac{1}{2}$ (T) and spin-1 (D) particles using the density matrix formulation, and three-dimensional (3D) spin transport hydrodynamics (STHD) simulation results of spin-polarized targets in the stagnation phase of indirectly driven ICF implosions. The depolarization of DT ions by strong self-generated magnetic fields can be captured by the spin transport equation. The neutron yield, angular distribution and neutron polarization can be obtained from the STHD simulations, which solve the hydrodynamic equations, magnetic induction equation, spin transport equations and fusion rate equation self-consistently. STHD simulations can be used to interpret the polarized ICF experiments and optimize the physics design of ICF polarized neutron source. We show as an example that an optimized spin alignment of the polarized DT fuel can reduce the neutron beam depolarization induced by polar mode asymmetries in indirectly driven implosions.

\section{Spin Transport Equation}
To obtain the spin transport equation for DT nuclei, we start from the single particle Schr\"odinger equation
\begin{equation}
 i\hbar\frac{\partial}{\partial t}\Psi_\alpha=\hat{H}\Psi_\alpha,
\end{equation}
with the Hamiltonian $\hat{H}$,
\begin{equation}
 \hat{H}=-\frac{\hbar^2}{2m}\nabla^2-\hat{\boldsymbol{\mu}}\cdot\boldsymbol{B},
\end{equation}
where $\alpha$ denotes the $\alpha$-th particle, $m$ is the particle mass, $\hat{\boldsymbol{\mu}}=\gamma\hat{\boldsymbol{s}}$ is the magnetic moment, $\gamma$ is the gyromagnetic ratio, $\hat{\boldsymbol{s}}$ is spin operator and $\boldsymbol{B}$ is the magnetic field. Here the spin-orbit and spin-spin interaction terms in the Hamiltonian are neglected because the interaction cross-sections are relatively small. The collisional depolarization caused by these interactions is estimated to be neglectable in ICF implosions \cite{more1983,goel1988}.
The wavefunction $\Psi_\alpha$ can be decomposed as \cite{marklund2007}
\begin{equation}
\label{eq_wavefunction}
 \Psi_\alpha=\sqrt{n_\alpha}\exp(iS_\alpha/\hbar)\varphi_\alpha,
\end{equation}
where $n_\alpha(\boldsymbol{r},t)$, $S_\alpha(\boldsymbol{r},t)$ are real functions. $\varphi_\alpha$ is the 2-spinor for spin-$\frac{1}{2}$ particle or 3-spinor for spin-1 particle. The single particle density $n_\alpha=\Psi_\alpha^\dagger\Psi_\alpha$ satisifies
\begin{equation}
\label{eq_density}
\begin{aligned}
 \frac{\partial n_\alpha}{\partial t} =& \left(\frac{\hat{H}\Psi_\alpha}{i\hbar}\right)^\dagger\Psi_\alpha+\Psi_\alpha^\dagger\left(\frac{\hat{H}\Psi_\alpha}{i\hbar}\right)\\
 =&\nabla\cdot\left[\left(\frac{i\hbar}{2m}\nabla\Psi_\alpha\right)^\dagger\Psi_\alpha+\Psi_\alpha^\dagger\left(\frac{i\hbar}{2m}\nabla\Psi_\alpha\right)\right]+\left[\left(\frac{-\hat{\boldsymbol{\mu}}\cdot\boldsymbol{B}}{i\hbar}\Psi_\alpha\right)^\dagger\Psi_\alpha+\Psi_\alpha^\dagger\left(\frac{-\hat{\boldsymbol{\mu}}\cdot\boldsymbol{B}}{i\hbar}\Psi_\alpha\right)\right].
\end{aligned}
\end{equation}
As the components of spin operator $\hat{\boldsymbol{s}}$ are Hermitian and $\hat{\boldsymbol{\mu}}\cdot\boldsymbol{B}$ is also Hermitian, the second term on the right-hand side of Eq. \ref{eq_density} is zero. The current density is defined as $\boldsymbol{J}_\alpha=\left(\frac{-i\hbar}{2m}\nabla\Psi_\alpha\right)^\dagger\Psi_\alpha+\Psi_\alpha^\dagger\left(\frac{-i\hbar}{2m}\nabla\Psi_\alpha\right)$, and then the equation of continuity can be obtained
\begin{equation}
 \frac{\partial n_\alpha}{\partial t} + \nabla\cdot \boldsymbol{J}_\alpha = 0.
\end{equation}
The velocity of the particle can be defined as \cite{marklund2007}
\begin{equation}
\label{eq_velocity}
 \boldsymbol{v}_\alpha=\boldsymbol{J}_\alpha/n_\alpha=\frac{\nabla S_\alpha-i\hbar\varphi_\alpha^\dagger\nabla\varphi_\alpha}{m}.
\end{equation}
The density matrix is defined as $\hat{\eta}_\alpha=\frac{\Psi_\alpha\otimes\Psi_\alpha^\dagger}{n_\alpha}$, where $\otimes$ denotes the tensor product. The evolution of density matrix can be written as
\begin{equation}
\begin{aligned}
 \frac{\partial n_\alpha\hat{\eta}_\alpha}{\partial t} + \nabla\cdot(n_\alpha\hat{\eta}_\alpha\boldsymbol{v}_\alpha) =& \frac{i}{\hbar}\left[\hat{\boldsymbol{\mu}}\cdot\boldsymbol{B},n_\alpha\hat{\eta}_\alpha\right]+\frac{i\hbar}{2m}\nabla\cdot\left[\nabla \hat{\eta}_\alpha,n_\alpha\hat{\eta}_\alpha\right],
\end{aligned}
\end{equation}
where $[\hat{A},\hat{B}]=\hat{A}\hat{B}-\hat{B}\hat{A}$ is the commutator.

The total density of a particle specie is $n = \sum _\alpha n_\alpha$. The density matrix of a particle specie is defined as $\hat{\eta}=\sum _\alpha n_\alpha \hat{\eta}_\alpha/n$ and the fluid velocity is $\boldsymbol{v}=\sum _\alpha n_\alpha \boldsymbol{v}_\alpha/n$. With these definitions, we can obtain the spin transport equation for a particle specie as
\begin{equation}
 \label{eq_ste}
 \frac{\partial n \hat{\eta}}{\partial t}+\nabla\cdot(n\hat{\eta}\boldsymbol{v})=\frac{i}{\hbar}\left[\hat{\boldsymbol{\mu}}\cdot\boldsymbol{B},n\hat{\eta}\right]-\nabla\cdot \hat{\boldsymbol{K}}+\nabla\cdot \hat{\boldsymbol{Q}},
\end{equation}
where $\hat{\boldsymbol{K}}=\sum_\alpha n_\alpha(\hat{\eta}_\alpha-\hat{\eta})(\boldsymbol{v}_\alpha-\boldsymbol{v})$ is the thermal-spin coupling, $\hat{\boldsymbol{Q}}=\frac{i\hbar}{2m}\sum_\alpha[\nabla \hat{\eta}_\alpha, n_\alpha \hat{\eta}_\alpha]$ is the nonlinear spin fluid contribution. The spin transport equation (\ref{eq_ste}) still contains
the explicit sum over all particles, and further statistical relations are needed to close the system. If the spin distribution and thermal distribution are not correlated, the thermal-spin coupling $\hat{\boldsymbol{K}}=0$. And if the typical fluid length scale $L\gg\lambda_{th}$, where $\lambda_{th}$ is the thermal de Broglie wavelength, the nonlinear spin fluid contribution $\hat{\boldsymbol{Q}}$ can be neglected.

The probability distribution for spin eigenstates can be obtained from diagonal terms of the density matrix. The trace of density matrix is conserved and unity $\rm{Tr}(\hat{\eta})=1$. For tritons, the probabilities for spin eigenstates $m_z=\{\frac{1}{2},-\frac{1}{2}\}$ are ${\eta}^T_{00}$ and ${\eta}^T_{11}$ respectively. For deuterons, the probabilities for spin eigenstates $m_z=\{1,0,-1\}$ are ${\eta}^D_{00}$, ${\eta}^D_{11}$ and ${\eta}^D_{22}$ respectively. The triton polarization in $+z$ direction is $p_z^T=\eta^T_{00}-\eta^T_{11}$. The vector polarization of deuteron is $p_z^D=\eta^D_{00}-\eta^D_{22}$ and the tensor polarization of deuteron is $p_{zz}^D=\eta^D_{00}-2\eta^D_{11}+\eta^D_{22}$. The fusion cross-section and neutron angular distributions can be obtained from probability distributions for spin eigenstates of DT. For simplicity, we adopt the formulas of Kulsrud \emph{et al.} \cite{kulsrud1982,kulsrud1986} in our simulations. The fusion cross-section of DT reaction can be calculated as
\begin{equation}
\begin{aligned}
 \label{fusion_cs}
 \sigma=&\sigma_0\left[\frac{3}{2}\left(\eta^T_{00}\eta^D_{00}+\eta^T_{11}\eta^D_{22}\right)+\eta^D_{11}+\frac{1}{2}\left(\eta^T_{00}\eta^D_{22}+\eta^T_{11}\eta^D_{00}\right)\right],
\end{aligned}
 \end{equation}
where $\sigma_0$ is the unpolarized cross-section. The total differential cross-section for neutrons is
\begin{equation}
\begin{aligned}
 \frac{d\sigma}{d\Omega}=&\frac{\sigma_0}{4\pi}\left[\frac{9}{4}\left(\eta^T_{00}\eta^D_{00}+\eta^T_{11}\eta^D_{22}\right)\sin^2\theta+\frac{1}{4}\left(\eta^T_{00}\eta^D_{22}+\eta^T_{11}\eta^D_{00}+2\eta^D_{11}\right)\left(3\cos^2\theta+1\right)\right],
\end{aligned}
\end{equation}
where $\theta$ is the polar angle. The differential cross-sections for neutrons with $m_z=\{\frac{1}{2},-\frac{1}{2}\}$ are $\frac{d\sigma^+}{d\Omega}$ and $\frac{d\sigma^-}{d\Omega}$ respectively, satisifying $\frac{d\sigma}{d\Omega}=\frac{d\sigma^+}{d\Omega}+\frac{d\sigma^-}{d\Omega}$ and
\begin{equation}
\label{neutron_diff}
\begin{aligned}
 &\frac{d\sigma^+}{d\Omega}-\frac{d\sigma^-}{d\Omega}=\frac{\sigma_0}{4\pi}\left[\frac{9}{4}\left(\eta^T_{00}-\eta^T_{11}\right)\left(\eta^D_{00}-2\eta^D_{11}+\eta^D_{22}\right)\sin^2\theta\cos^2\theta\right.\\
 &\left.-\frac{9}{4}\left(\eta^T_{00}\eta^D_{00}-\eta^T_{11}\eta^D_{22}\right)\sin^4\theta+\frac{1}{4}\left(2\eta^T_{00}\eta^D_{11}-2\eta^T_{11}\eta^D_{11}+\eta^T_{11}\eta^D_{00}-\eta^T_{00}\eta^D_{22}\right)\left(3\cos^2\theta-1\right)^2\right],
\end{aligned}
\end{equation}
The cross-section and differential cross-sections (\ref{fusion_cs})-(\ref{neutron_diff}) are used in the fusion rate equations to obtain the neutron yield, neutron angular distribution and neutron polarization of polarized DT fusion.

\section{Numerical algorithm of SPINSIM}

As the spin transport equation contains fluid quantities $n$, $\boldsymbol{v}$ and magnetic field $\boldsymbol{B}$, it must be solved in combination with hydrodynamic equations and magnetic induction equation. A numerical scheme to solve the spin transport equation is developed and implementd in a 3D STHD simulation code SPINSIM, which numerically guarantees that the diagonal terms of the density matrix are bounded in [0,1] and $\rm{Tr}(\hat{\eta})=1$.  The STHD simulation code SPINSIM solves the hydrodynamics equations, magnetic induction equation, spin transport equation and fusion rate equation, 
\begin{equation}
\label{S1}
\frac{\partial \rho}{\partial t} + \nabla\cdot(\rho \boldsymbol{v}) = \left(\frac{\partial \rho}{\partial t}\right)_{fusion},
\end{equation}

\begin{equation}
\label{S2}
\frac{\partial \rho \boldsymbol{v}}{\partial t} + \nabla\cdot(\rho \boldsymbol{v}\otimes\boldsymbol{v}) + \nabla p = \boldsymbol{v}\left(\frac{\partial \rho}{\partial t}\right)_{fusion},
\end{equation}

\begin{equation}
\label{S3}
\frac{\partial \rho E}{\partial t} + \nabla\cdot[(\rho E +p)\boldsymbol{v}] = E\left(\frac{\partial \rho}{\partial t}\right)_{fusion},
\end{equation}

\begin{equation}
\label{S4}
\frac{\partial \boldsymbol{B}}{\partial t}-\nabla\times(\boldsymbol{v}\times\boldsymbol{B})=\nabla\times\left(\frac{\nabla p_e}{n_e e}\right),
\end{equation}

\begin{equation}
\label{S5}
\frac{\partial \rho f}{\partial t}+\nabla\cdot(\rho f\boldsymbol{v})=\left(\frac{\partial \rho f}{\partial t}\right)_{fusion},
\end{equation}

\begin{equation}
\label{S6}
\frac{\partial \rho f\hat{\eta}}{\partial t}+\nabla\cdot(\rho f\hat{\eta}\boldsymbol{v})=\frac{i}{\hbar}\left[\hat{\boldsymbol{\mu}}\cdot\boldsymbol{B},\rho f\hat{\eta}\right]+\hat{\eta}\left(\frac{\partial \rho f}{\partial t}\right)_{fusion},
\end{equation}

\begin{equation}
\label{S7}
\left(\frac{\partial \rho}{\partial t}\right)_{fusion}=\left(\frac{\partial \rho f}{\partial t}\right)_{fusion}=-\frac{\left<\sigma v\right>}{m_D+m_T}(\rho f)^2.
\end{equation}
Eqs. \eqref{S1}-\eqref{S3} are hydrodynamics equations, where $\rho$ is mass density, $\boldsymbol{v}$ is fluid velocity, $p$ is pressure, $E=v^2/2+E_{int}$ is total energy per unit mass, $E_{int}$ is the internal energy per unit mass. Eq. \eqref{S4} is the magnetic induction equation including magnetic transport by bulk advection term and Biermann battery source term, where $\boldsymbol{B}$ is magnetic field, $p_e$ is electron pressure, $n_e$ is electron number density, $e$ is elementary charge. The Lorentz force on the fluid flow is neglected here because the ratio of thermal pressure to magnetic pressure $\beta=\frac{p}{\boldsymbol{B}^2/2\mu_0}\gg1$ for self-generated magnetic fields in ICF plasma ($\mu_0$ is the vacuum permeability). Eq. \eqref{S5} is the continuity equation for deuterium-tritium (DT) fuels, where $f$ is the mass fraction of DT fuel. Eq. \eqref{S6} is the spin transport equation for deuteron (D) and triton (T), where $\hat{\eta}$ is the density matrix, $\hat{\boldsymbol{\mu}}$ is magnetic moment. Eq. \eqref{S7} is the fusion rate equation for equimolar DT fuels, $\left<\sigma v\right>$ is the fusion reactivity, $m_D$ and $m_T$ are masses of DT respectively. Alpha particle energy deposition is neglected because the plasma conditions investigated here are lower than the self-heating limit \cite{hurricane2016}. Eqs. \eqref{S1}-\eqref{S6} can be written in the form of conservation laws, 
\begin{equation}
\label{S8}
\frac{\partial U}{\partial t} + \frac{\partial F(U)}{\partial x} + \frac{\partial G(U)}{\partial y} + \frac{\partial H(U)}{\partial z} = S(U),
\end{equation}
where $U$ is the conserved variable, ($F$, $G$, $H$) are components of the flux, $S$ is the source. Eq. \eqref{S8} can be solved using the operator splitting method. In each time-step, the conserved variables are first updated by
\begin{equation}
\label{S9}
\frac{\partial U}{\partial t} + \frac{\partial F(U)}{\partial x} + \frac{\partial G(U)}{\partial y} + \frac{\partial H(U)}{\partial z} =0,
\end{equation}
and then updated by
\begin{equation}
\label{S10}
\frac{\partial U}{\partial t} = S(U),
\end{equation}
Eq. \eqref{S9} is solved with the finite volume method on 3D Cartesian grid, 
\begin{equation}
\label{S11}
U^{n+1/2}_{i,j,k}=U^{n}_{i,j,k}-\Delta t\left(\frac{F^{n}_{i+1/2,j,k}-F^{n}_{i-1/2,j,k}}{\Delta x}+\frac{G^{n}_{i,j+1/2,k}-G^{n}_{i,j-1/2,k}}{\Delta y}+\frac{H^{n}_{i,j,k+1/2}-H^{n}_{i,j,k-1/2}}{\Delta z}\right),
\end{equation}
where $i, j, k$ refer to the discrete cell index in $x, y, z$ directions respectively, $U^n$ is the state at $t=t^n$, $F^{n}_{i\pm1/2,j,k}$, $G^{n}_{i,j\pm1/2,k}$, $H^{n}_{i,j,k\pm1/2}$ are the fluxes at the cell interfaces in x, y and z directions respectively, $U^{n+1/2}$ is the intermediate state used as the input of Eq. \eqref{S10} to obtain the state $U^{n+1}$ at $t=t^n+\Delta t$.

 For hydrodynamics equations \eqref{S1}-\eqref{S3} and magnetic induction equation \eqref{S4}, the conserved variables are
 \begin{equation}
 \label{S12}
 U = \left(\rho, \rho v_x, \rho v_y, \rho v_z, \rho E, B_x, B_y, B_z\right)^T,
 \end{equation}
and the fluxes are given by
\begin{equation}
 \label{S13}
 \begin{aligned}
 &F = \left(\rho v_x, \rho v_x^2+p, \rho v_x v_y, \rho v_x v_z, (\rho E+p)v_x, 0, v_x B_y-v_y B_x, v_x B_z-v_z B_x\right)^T,\\
 &G = \left(\rho v_y, \rho v_x v_y, \rho v_y^2 +p, \rho v_y v_z, (\rho E+p)v_y, v_y B_x-v_x B_y, 0, v_y B_z-v_z B_y\right)^T,\\
 &H = \left(\rho v_z, \rho v_x v_z, \rho v_y v_z, \rho v_z^2 +p, (\rho E+p)v_z, v_z B_x-v_x B_z, v_z B_y-v_y B_z, 0\right)^T.
 \end{aligned}
 \end{equation}
 These fluxes at the cell interfaces can be calculated using Riemann solvers. The Harten-Lax-van Leer (HLL) approximate Riemann solver \cite{toro2009} is adopted in SPINSIM. For example, the flux $F^{n}_{i-1/2,j,k}$ can be calculated as 
\begin{equation}
\label{S14}
F^{n}_{i-1/2,j,k}=\frac{\alpha^+ F(U^L_{i-1/2,j,k})+\alpha^-F(U^R_{i-1/2,j,k})-\alpha^+\alpha^-(U^R_{i-1/2,j,k}-U^L_{i-1/2,j,k})}{\alpha^++\alpha^-},
\end{equation}
where $U^L_{i-1/2,j,k}$ and $U^R_{i-1/2,j,k}$ are left and right state at the cell interface, which can be reconstructed using the piecewise linear method as
\begin{equation}
 \label{S15}
 \begin{aligned}
 &U^L_{i-1/2,j,k}=U^n_{i-1,j,k} + 0.5\textrm{minmod}(U^n_{i-1,j,k}-U^n_{i-2,j,k},0.5(U^n_{i,j,k}-U^n_{i-2,j,k}), U^n_{i,j,k}-U^n_{i-1,j,k}),\\
 &U^R_{i-1/2,j,k}=U^n_{i,j,k} - 0.5\textrm{minmod}(U^n_{i,j,k}-U^n_{i-1,j,k},0.5(U^n_{i+1,j,k}-U^n_{i-1,j,k}), U^n_{i+1,j,k}-U^n_{i,j,k}).
 \end{aligned}
 \end{equation}
 The minmod slope limiter is given as
 \begin{equation}
 \label{S16}
 \textrm{minmod}(a, b, c) =0.25\left|\textrm{sign}(a)+\textrm{sign}(b)\right|(\textrm{sign}(a)+\textrm{sign}(c))\textrm{min}(|a|, |b|, |c|).
 \end{equation}
 $\alpha^+=\textrm{max}(0, v_x^L+c_s^L, v_x^R+c_s^R)$ and $\alpha^-=-\textrm{min}(0,v_x^L-c_s^L,v_x^R-c_s^R)$ are the maximum right and left propagating wave speed respectively, where $c_s$ is the sound speed. The time step $\Delta t$ should satisfy the Courant-Friedrichs-Levy condition, $\textrm{max}(\alpha^+_x,\alpha^-_x)\frac{\Delta t}{\Delta x}+\textrm{max}(\alpha^+_y,\alpha^-_y)\frac{\Delta t}{\Delta y}+\textrm{max}(\alpha^+_z,\alpha^-_z)\frac{\Delta t}{\Delta z}\le1$.

For the continuity equation of DT fuel \eqref{S5} and spin transport equation \eqref{S6}, the conserved variables are
\begin{equation}
 \label{S17}
 U = \left(\rho f, \rho f \hat{\eta}\right)^T,
 \end{equation}
and the fluxes are given by
\begin{equation}
 \label{S18}
 \begin{aligned}
 &F = \left(\rho f v_x, \rho f \hat{\eta} v_x \right)^T,\\
 &G = \left(\rho f v_y, \rho f \hat{\eta} v_y \right)^T,\\
 &H = \left(\rho f v_z, \rho f \hat{\eta} v_z \right)^T.
 \end{aligned}
 \end{equation}
 These fluxes are calculated using the fluxes of Eq. \eqref{S1} as
 \begin{equation}
 \label{S19}
 \begin{aligned}
 (\rho f v_x)_{i-1/2,j,k} = &
 \begin{cases}
 (\rho v_x)^{HLL}_{i-1/2,j,k} f_{i-1,j,k} & \text{if } (\rho v_x)^{HLL}_{i-1/2,j,k}\ge0,\\
 (\rho v_x)^{HLL}_{i-1/2,j,k} f_{i,j,k} & \text{if } (\rho v_x)^{HLL}_{i-1/2,j,k}<0,
 \end{cases}\\
 (\rho f \hat{\eta} v_x)_{i-1/2,j,k} = &
 \begin{cases}
 (\rho v_x)^{HLL}_{i-1/2,j,k} f_{i-1,j,k} \hat{\eta}_{i-1,j,k} & \text{if } (\rho v_x)^{HLL}_{i-1/2,j,k}\ge0,\\
 (\rho v_x)^{HLL}_{i-1/2,j,k} f_{i,j,k} \hat{\eta}_{i,j,k} & \text{if } (\rho v_x)^{HLL}_{i-1/2,j,k}<0.
 \end{cases}
 \end{aligned}
 \end{equation}
 This flux calculation method numerically guarantees that $f$ and the diagonal terms of $\hat{\eta}$ are bounded in $[0,1]$, and the trace of $\hat{\eta}$ is unity.
 
 The magnetic field generation from the Biermann battery effect is calculated by
 \begin{equation}
 \label{S20}
 \begin{aligned}
 \left(B_x\right)^{n+1}_{i,j,k}=\left(B_x\right)^{n+1/2}_{i,j,k}+\frac{\Delta t}{4e\Delta y \Delta z}\left(\frac{(p_e)^{n+1/2}_{i,j+1,k+1}-(p_e)^{n+1/2}_{i,j+1,k-1}}{(n_e)^{n+1/2}_{i,j+1,k}}-\frac{(p_e)^{n+1/2}_{i,j-1,k+1}-(p_e)^{n+1/2}_{i,j-1,k-1}}{(n_e)^{n+1/2}_{i,j-1,k}}\right.\\
 \left.-\frac{(p_e)^{n+1/2}_{i,j+1,k+1}-(p_e)^{n+1/2}_{i,j-1,k+1}}{(n_e)^{n+1/2}_{i,j,k+1}}+\frac{(p_e)^{n+1/2}_{i,j+1,k-1}-(p_e)^{n+1/2}_{i,j-1,k-1}}{(n_e)^{n+1/2}_{i,j,k-1}}\right),\\
 \left(B_y\right)^{n+1}_{i,j,k}=\left(B_y\right)^{n+1/2}_{i,j,k}+\frac{\Delta t}{4e\Delta x \Delta z}\left(\frac{(p_e)^{n+1/2}_{i+1,j,k+1}-(p_e)^{n+1/2}_{i-1,j,k+1}}{(n_e)^{n+1/2}_{i,j,k+1}}-\frac{(p_e)^{n+1/2}_{i+1,j,k-1}-(p_e)^{n+1/2}_{i-1,j,k-1}}{(n_e)^{n+1/2}_{i,j,k-1}}\right.\\
 \left.-\frac{(p_e)^{n+1/2}_{i+1,j,k+1}-(p_e)^{n+1/2}_{i+1,j,k-1}}{(n_e)^{n+1/2}_{i+1,j,k}}+\frac{(p_e)^{n+1/2}_{i-1,j,k+1}-(p_e)^{n+1/2}_{i-1,j,k-1}}{(n_e)^{n+1/2}_{i-1,j,k}}\right),\\
 \left(B_z\right)^{n+1}_{i,j,k}=\left(B_z\right)^{n+1/2}_{i,j,k}+\frac{\Delta t}{4e\Delta x \Delta y}\left(\frac{(p_e)^{n+1/2}_{i+1,j+1,k}-(p_e)^{n+1/2}_{i+1,j-1,k}}{(n_e)^{n+1/2}_{i+1,j,k}}-\frac{(p_e)^{n+1/2}_{i-1,j+1,k}-(p_e)^{n+1/2}_{i-1,j-1,k}}{(n_e)^{n+1/2}_{i-1,j,k}}\right.\\
 \left.-\frac{(p_e)^{n+1/2}_{i+1,j+1,k}-(p_e)^{n+1/2}_{i-1,j+1,k}}{(n_e)^{n+1/2}_{i,j+1,k}}+\frac{(p_e)^{n+1/2}_{i+1,j-1,k}-(p_e)^{n+1/2}_{i-1,j-1,k}}{(n_e)^{n+1/2}_{i,j-1,k}}\right).
 \end{aligned}
 \end{equation} 
 
The equation of state (EOS) used in SPINSIM includes contributions from ions, electrons and radiation. The temperatures of ions, electrons and radiation are assumed to be in equilibrium, \textit{i.e.} $T_i=T_e=T_r=T$. The pressure and internal energy are calculated as the sum over all components \cite{drake2018}
\begin{equation}
\label{S21}
\begin{aligned}
p&=p_i+p_e+p_r,\\
E_{int} &= E_i+E_e+E_r.
\end{aligned}
\end{equation}
The ions and electrons are treated as ideal gas with constant adiabatic index
\begin{equation}
\label{S22}
\begin{aligned}
p_i+p_e=\frac{\rho f}{m_i^{fuel}}\left(1+Z_i^{fuel}\right)k_BT&+\frac{\rho (1-f)}{m_i^{shell}}\left(1+Z_i^{shell}\right)k_BT,\\
E_i+E_e=&\frac{p_i+p_e}{(\gamma-1)\rho}.
\end{aligned}
\end{equation}
$m_i^{fuel}$ and $m_i^{shell}$ are the average ion masses for DT fuel and shell material respectively, $Z_i^{fuel}$ and $Z_i^{shell}$ are the average ionization degrees for DT fuel and shell material respectively, $k_B$ is the Boltzmann constant, $\gamma$ is the adiabatic index. $Z_i^{fuel}=1$ and $Z_i^{shell}=6$ are used for the DT fuel and high density carbon (HDC) shell. The adiabatic index is taken as $\gamma=5/3$. The radiation is assumed to be blackbody radiation in local thermodynamic equilibrium,
\begin{equation}
\label{S23}
p_r = \frac{4\sigma_B T^4}{3 c},\ E_r = \frac{3 p_r}{\rho},
\end{equation}
where $\sigma_B$ is the Stephan-Boltzmann constant, $c$ is the speed of light.

The fusion reaction equation \eqref{S7} is solved with the analytical solution
\begin{equation}
\label{S24}
\begin{aligned}
\rho^{n+1}_{i,j,k} f^{n+1}_{i,j,k}=\frac{\rho^{n+1/2}_{i,j,k}f^{n+1/2}_{i,j,k}}{1+\frac{\rho^{n+1/2}_{i,j,k}f^{n+1/2}_{i,j,k}}{m_D+m_T}\left<\sigma v\right>^{n+1/2}_{i,j,k}\Delta t},\\
\rho^{n+1}_{i,j,k} = \rho^{n+1}_{i,j,k} f^{n+1}_{i,j,k} + \rho^{n+1/2}_{i,j,k}(1-f^{n+1/2}_{i,j,k}).
\end{aligned}
\end{equation}
The fusion reactivity $\left<\sigma v\right>^{n+1/2}_{i,j,k}$ can be calculated as \cite{kulsrud1982,*kulsrud1986}
\begin{equation}
\label{S25}
\begin{aligned}
\left<\sigma v\right>^{n+1/2}_{i,j,k}=\delta\left<\sigma_0 v\right>^{n+1/2}_{i,j,k},&\\
\delta=\frac{3}{2}\left(\left(\eta^T_{00}\right)^{n+1/2}_{i,j,k}\left(\eta^D_{00}\right)^{n+1/2}_{i,j,k}+\left(\eta^T_{11}\right)^{n+1/2}_{i,j,k}\right.&\left.\left(\eta^D_{22}\right)^{n+1/2}_{i,j,k}\right)+\left(\eta^D_{11}\right)^{n+1/2}_{i,j,k}\\
+\frac{1}{2}\left(\left(\eta^T_{00}\right)^{n+1/2}_{i,j,k}\left(\eta^D_{22}\right)^{n+1/2}_{i,j,k}+\left(\eta^T_{11}\right)^{n+1/2}_{i,j,k}\right.&\left.\left(\eta^D_{00}\right)^{n+1/2}_{i,j,k}\right),
\end{aligned}
\end{equation}
where $\left<\sigma_0 v\right>$ is the unpolarized reactivity which can be calculated from the ion temperature by the empirical formula of Bosch and Hale \cite{bosch1992}.
The neutron yield in each cell is calculated by
\begin{equation}
\label{S26}
N^{n+1}_{i,j,k}=N^{n}_{i,j,k}+\frac{\rho^{n+1/2}_{i,j,k}f^{n+1/2}_{i,j,k}-\rho^{n+1}_{i,j,k} f^{n+1}_{i,j,k}}{m_D+m_T}\Delta x\Delta y\Delta z.
\end{equation}
The angular distribution of neutron is
\begin{equation}
\label{S27}
\begin{aligned}
\frac{dN}{d\Omega}&=\frac{dN^+}{d\Omega}+\frac{dN^-}{d\Omega}=C_1\sin^2\theta+C_2\left(3\cos^2\theta+1\right),\\
\frac{dN^+}{d\Omega}-\frac{dN^-}{d\Omega}&=C_3\sin^2\theta \cos^2\theta+C_4\sin^4\theta+C_5\left(3\cos^2\theta-1\right)^2,
\end{aligned}
\end{equation}
where $\frac{dN}{d\Omega}$ is the total angular distribution, $\frac{dN^\pm}{d\Omega}$ are the angular distribution for $m_z^n=\pm\frac{1}{2}$ state respectively, the coefficients are
\begin{equation}
 \label{S28}
\begin{aligned}
(C_1)^{n+1}_{i,j,k}&=(C_1)^n_{i,j,k}+2.25\frac{N^{n+1}_{i,j,k}-N^{n}_{i,j,k}}{4\pi\delta}\left(\left(\eta_{00}^T\right)^{n+1/2}_{i,j,k}\left(\eta_{00}^D\right)^{n+1/2}_{i,j,k}+\left(\eta_{11}^T\right)^{n+1/2}_{i,j,k}\left(\eta_{22}^D\right)^{n+1/2}_{i,j,k}\right),\\
(C_2)^{n+1}_{i,j,k}&=(C_2)^n_{i,j,k}+0.25\frac{N^{n+1}_{i,j,k}-N^{n}_{i,j,k}}{4\pi\delta}\left(\left(\eta_{00}^T\right)^{n+1/2}_{i,j,k}\left(\eta_{22}^D\right)^{n+1/2}_{i,j,k}+\left(\eta_{11}^T\right)^{n+1/2}_{i,j,k}\left(\eta_{00}^D\right)^{n+1/2}_{i,j,k}+2\left(\eta_{11}^D\right)^{n+1/2}_{i,j,k}\right),\\
(C_3)^{n+1}_{i,j,k}&=(C_3)^n_{i,j,k}+2.25\frac{N^{n+1}_{i,j,k}-N^{n}_{i,j,k}}{4\pi\delta}\left(\left(\eta_{00}^T\right)^{n+1/2}_{i,j,k}-\left(\eta_{11}^T\right)^{n+1/2}_{i,j,k}\right)\left(\left(\eta_{00}^D\right)^{n+1/2}_{i,j,k}+\left(\eta_{22}^D\right)^{n+1/2}_{i,j,k}-2\left(\eta_{11}^D\right)^{n+1/2}_{i,j,k}\right),\\
(C_4)^{n+1}_{i,j,k}&=(C_4)^n_{i,j,k}+2.25\frac{N^{n+1}_{i,j,k}-N^{n}_{i,j,k}}{4\pi\delta}\left(\left(\eta_{11}^T\right)^{n+1/2}_{i,j,k}\left(\eta_{22}^D\right)^{n+1/2}_{i,j,k}-\left(\eta_{00}^T\right)^{n+1/2}_{i,j,k}\left(\eta_{00}^D\right)^{n+1/2}_{i,j,k}\right),\\
(C_5)^{n+1}_{i,j,k}&=(C_5)^n_{i,j,k}+0.25\frac{N^{n+1}_{i,j,k}-N^{n}_{i,j,k}}{4\pi\delta}\\
&\left(2\left(\eta^T_{00}\right)^{n+1/2}_{i,j,k}\left(\eta^D_{11}\right)^{n+1/2}_{i,j,k}-2\left(\eta^T_{11}\right)^{n+1/2}_{i,j,k}\left(\eta^D_{11}\right)^{n+1/2}_{i,j,k}+\left(\eta^T_{11}\right)^{n+1/2}_{i,j,k}\left(\eta^D_{00}\right)^{n+1/2}_{i,j,k}-\left(\eta^T_{00}\right)^{n+1/2}_{i,j,k}\left(\eta^D_{22}\right)^{n+1/2}_{i,j,k}\right).
\end{aligned}
\end{equation}
The evolution of the density matrix in magnetic fields follows the von Neumann equation
\begin{equation}
\label{S29}
\frac{\partial \hat{\eta}}{\partial t}=\frac{i}{\hbar}\left[\hat{\boldsymbol{\mu}}\cdot\boldsymbol{B},\hat{\eta}\right],
\end{equation}
where $\hat{\boldsymbol{\mu}}=\gamma\hat{\boldsymbol{s}}$ is the magnetic moment, $\gamma$ is the gyromagnetic ratio, $\hat{\boldsymbol{s}}$ is spin operator. For spin-$\frac{1}{2}$ particles, the components of $\hat{\boldsymbol{s}}$ in $z$-basis are
\begin{equation}
\label{S30}
 \hat{s}_x=\frac{\hbar}{2}
     \begin{pmatrix}
       0 & 1\\
       1 & 0
     \end{pmatrix},
 \hat{s}_y=\frac{\hbar}{2}
     \begin{pmatrix}
       0 & -i\\
       i & 0
     \end{pmatrix},
 \hat{s}_z=\frac{\hbar}{2}
     \begin{pmatrix}
       1 & 0\\
       0 & -1
     \end{pmatrix}.
\end{equation}
For spin-1 particles, the components of $\hat{\boldsymbol{s}}$ are
\begin{equation}
\label{S31}
  \hat{s}_x=\frac{\hbar}{\sqrt{2}}
     \begin{pmatrix}
       0 & 1 & 0\\
       1 & 0 & 1\\
       0 & 1 & 0
     \end{pmatrix},
 \hat{s}_y=\frac{\hbar}{\sqrt{2}}
     \begin{pmatrix}
       0 & -i & 0\\
       i & 0  & -i\\
       0 & i & 0
     \end{pmatrix},
 \hat{s}_z=\hbar
     \begin{pmatrix}
       1 & 0 & 0\\
       0 & 0 & 0\\
       0 & 0 & -1
     \end{pmatrix}.
\end{equation}
Eq. \eqref{S29} is solved with the analytical solution,
\begin{equation}
\label{S32}
\hat{\eta}^{n+1}_{i,j,k} = \hat{U}\hat{\eta}^{n+1/2}_{i,j,k}\hat{U}^\dagger,
\end{equation}
where $\hat{U}=\exp\left(\frac{i}{\hbar}\hat{\boldsymbol{\mu}}\cdot\boldsymbol{B}^{n+1/2}_{i,j,k}\Delta t\right)$ is the propagator. The propagator $\hat{U}$ for spin-$\frac{1}{2}$ particles can be written as
\begin{equation}
\label{S33}
 \hat{U}=
     \begin{pmatrix}
       \cos(\phi)+i\frac{B_z}{|\boldsymbol{B}|}\sin(\phi) & \frac{iB_x+B_y}{|\boldsymbol{B}|}\sin(\phi)\\
       \frac{iB_x-B_y}{|\boldsymbol{B}|}\sin(\phi) & \cos(\phi)-i\frac{B_z}{|\boldsymbol{B}|}\sin(\phi)
     \end{pmatrix},
\end{equation}
where $\phi=\frac{\gamma|\boldsymbol{B}|\Delta t}{2}$. The propagator $\hat{U}$ for spin-1 particles can be calculated as
\begin{equation}
\label{S34}
\begin{aligned}
 \hat{U}=&\left[\hat{I}-\left(\frac{\hat{s}_x}{\hbar}\right)^2\left(1-\cos(\phi_x)\right)+i\left(\frac{\hat{s}_x}{\hbar}\right)\sin(\phi_x)\right]\\
                &\cdot\left[\hat{I}-\left(\frac{\hat{s}_y}{\hbar}\right)^2\left(1-\cos(\phi_y)\right)+i\left(\frac{\hat{s}_y}{\hbar}\right)\sin(\phi_y)\right]\\
                &\cdot\left[\hat{I}-\left(\frac{\hat{s}_z}{\hbar}\right)^2\left(1-\cos(\phi_z)\right)+i\left(\frac{\hat{s}_z}{\hbar}\right)\sin(\phi_z)\right],
\end{aligned}
\end{equation}
where $\hat{I}$ is the identity matrix and $\phi_{\{x,y,z\}}=\gamma B_{\{x,y,z\}}\Delta t$. The analytical solution \eqref{S32} ensures that the diagonal terms of $\hat{\eta}$ are bounded in $[0,1]$, and the trace of $\hat{\eta}$ is unity.

\begin{figure}[htbp]
\includegraphics[width=0.7\textwidth]{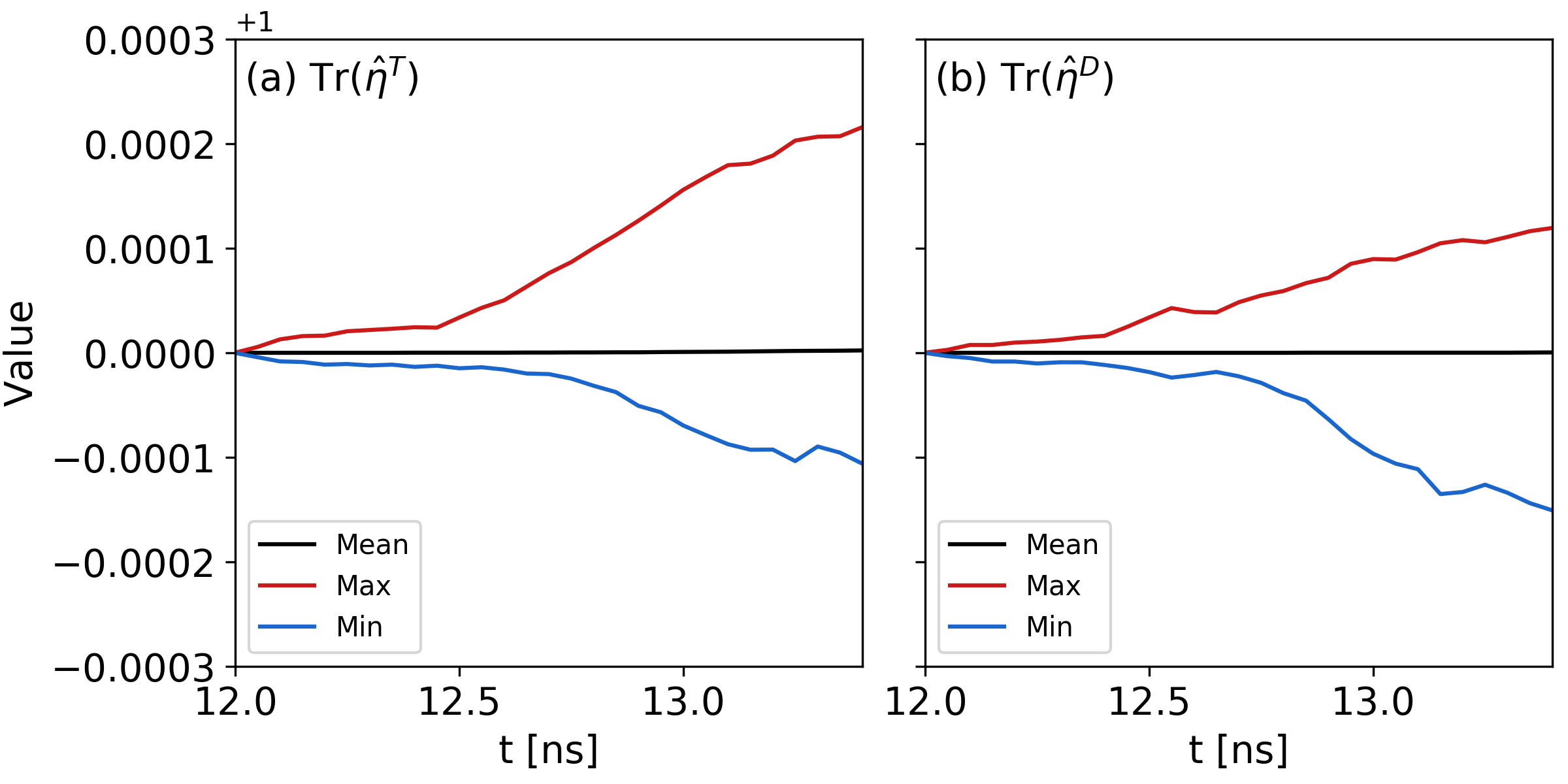}
\centering
\caption{\label{sup_fig_1} The simulation results for the traces of density matrices of (a) triton and (b) deuteron. The black lines are the average values over all computation cells, the red lines are maximum cell values and the blue lines are minimum cell values.}
\end{figure}

The numerical algorithm of SPINSIM can be efficiently parallelized on Graphics Processing Unit (GPU). Fig. \ref{sup_fig_1} shows the simulation results for the traces of density matrices of triton and deuteron. The simulations are carried out on GPUs using single precision float numbers. The numerical algorithm is stable and accurate as the maximum errors for the traces of density matrices remain within $\pm$0.03\% during the simulations. 

\section{Simulation parameters and results}

\begin{figure}[htbp]
\includegraphics[width=0.7\textwidth]{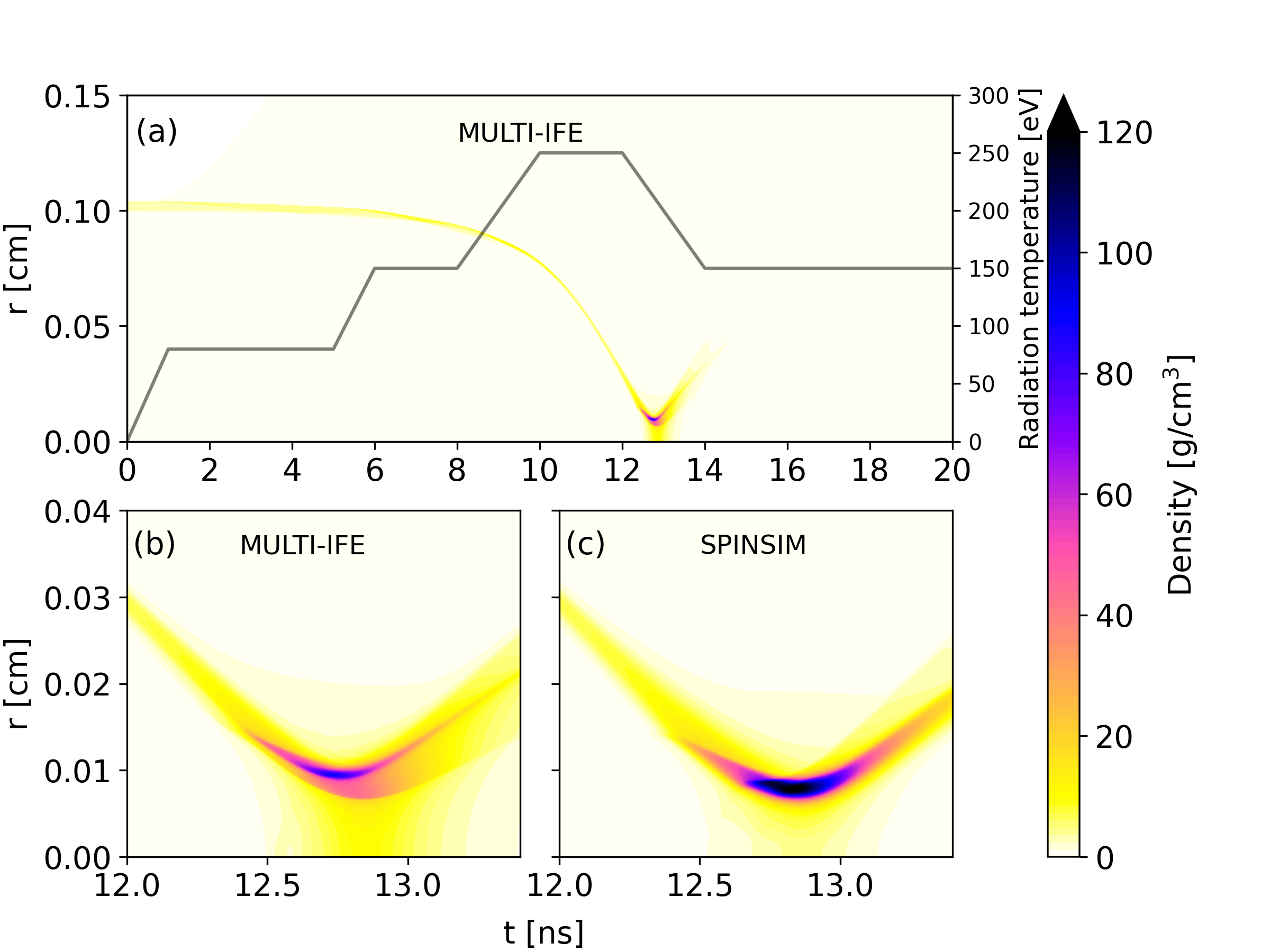}
\centering
\caption{\label{sup_fig_2} (a) Target mass density evolution from MULTI-IFE simulations. The black solid line is the radiation profile used to drive the implosion. (b) MULTI-IFE simulation results of the stagnation phase. (c) SPINSIM simulation results of the stagnation phase. }
\end{figure}

\begin{figure}[htbp]
\includegraphics[width=0.7\textwidth]{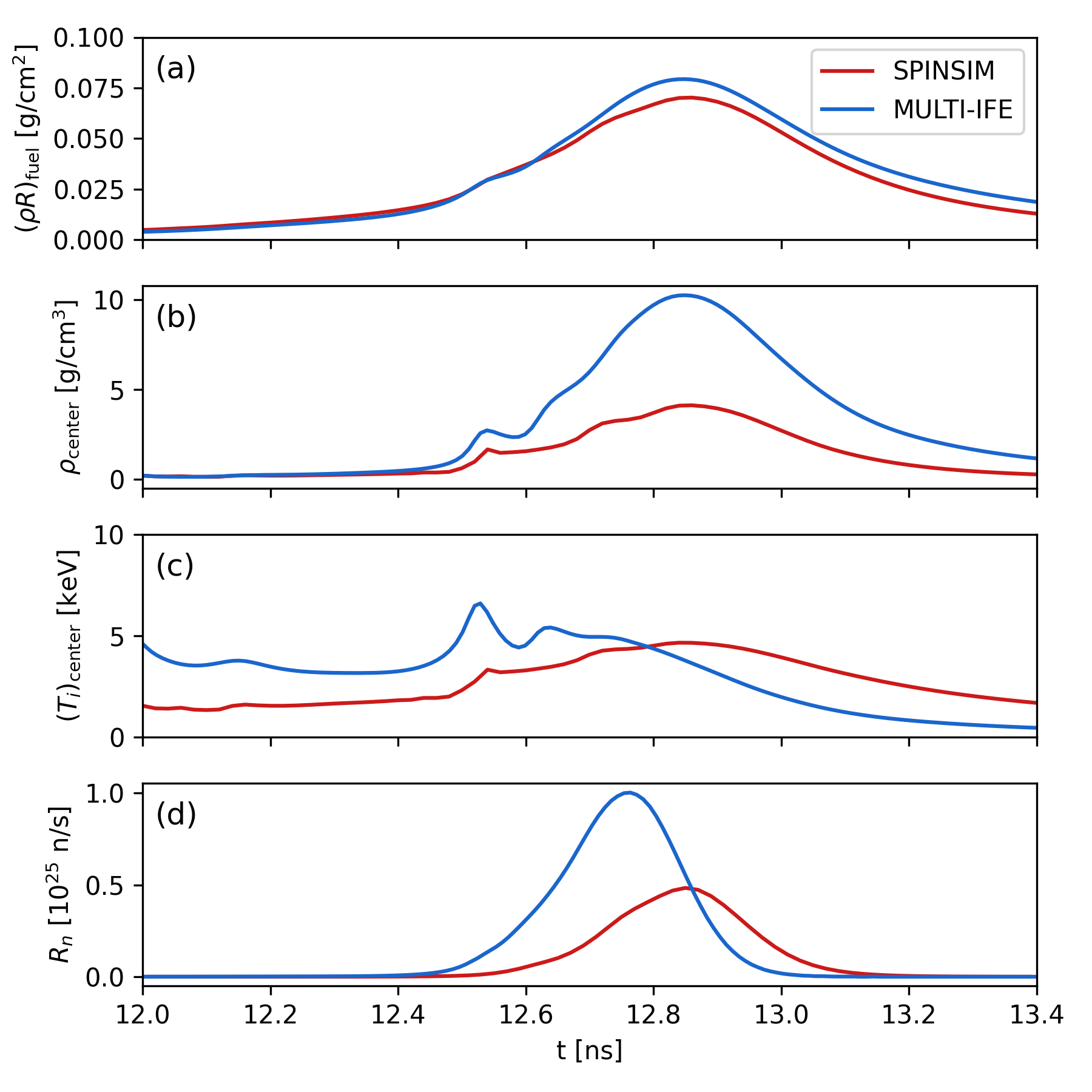}
\centering
\caption{\label{sup_fig_3} Simulation results of polarized target implosion of SPINSIM and MULTI-IFE. (a) Fuel areal density $(\rho R)_{fuel}$, (b) fuel center density $\rho_{center}$, (c) fuel center ion temperature $(T_i)_{center}$, (d) neutron production rate $R_n$.}
\end{figure}

\begin{figure}[htbp]
\includegraphics[width=0.5\textwidth]{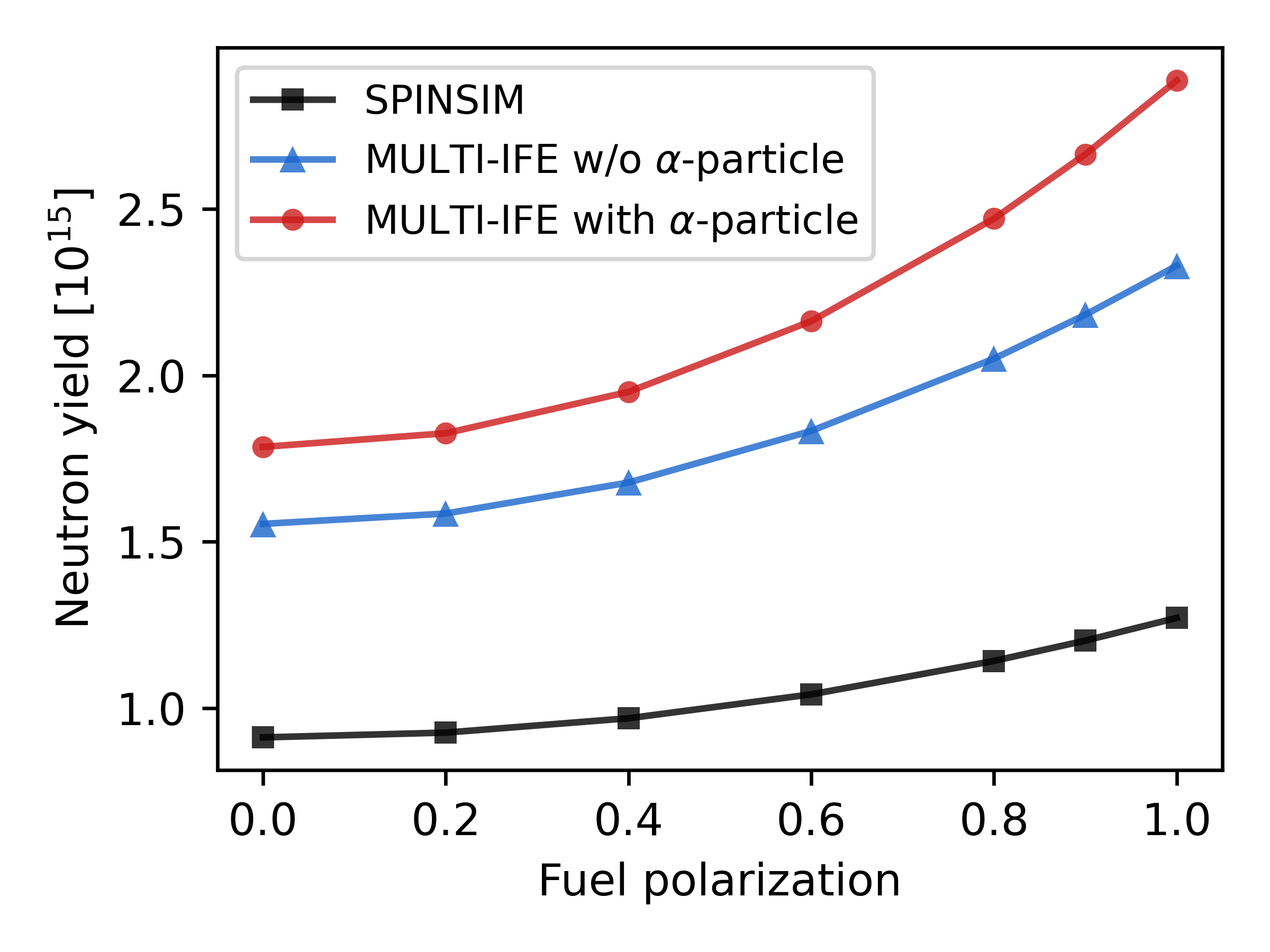}
\centering
\caption{\label{sup_fig_4} Neutron yields as function of the initial fuel polarization ($p_z^T=p_z^D=p_{zz}^D$ is assumed) from SPINSIM and MULTI-IFE simulations. }
\end{figure}

MULTI-IFE \cite{ramis2016} is used to generate initial conditions for SPINSIM. The radiation driven implosion of polarized target with HDC shell is simulated with MULTI-IFE, and the results of the whole implosion process are shown in Fig. \ref{sup_fig_2}(a). The hydrodynamic quantities $\rho$, $\boldsymbol{v}$ and $p$ at 12.0 ns from MULTI-IFE results are used as initial conditions for SPINSIM simulations. The simulation results of the stagnation phase of the implosion without perturbations from MULTI-IFE and SPINSIM are shown in Fig. \ref{sup_fig_2}(b) and (c). The 3D SPINSIM simulation is carried out on 400$\times$400$\times$400 uniform grid cells, the simulation domain size is 0.08$\times$0.08$\times$0.08 cm$^3$. As SPINSIM uses Eulerian finite-volume method, numerical diffusion can not be avoided and can lead to the mixing of shell and fuel materials. The hot-spot density in the stagnation phase for SPINSIM is lower than MULTI-IFE, which is a Lagrangian code without numerical diffusion, as shown in Fig. \ref{sup_fig_2}(b) and (c). Some fuel is mixed into the shell and the shell density in SPINSIM is higher than in MULTI-IFE. Fig. \ref{sup_fig_3} shows the temporal evolution of fuel areal density $(\rho R)_{fuel}$ (a), fuel center density $\rho_{center}$ (b), fuel center ion temperature $(T_i)_{center}$ (c) and neutron production rate $R_n$ (d). The fuel areal density and ion temperature from SPINSIM and MULTI-IFE simulations are close, but the smaller fuel density in SPINSIM leads to smaller neutron production rate and neutron yield as shown in Fig. \ref{sup_fig_3}, \ref{sup_fig_4} and Table \ref{sup_tab_1}. The alpha-particle self-heating effect is neglected in SPINSIM because the fuel areal density $\rho R$ is smaller than 0.1 g/cm$^2$ and the yield amplifications due to self-heating $Y_\alpha/Y_{\text{no-}\alpha}$ from MULTI-IFE simulations are smaller than 1.3 \cite{hurricane2016} as shown in Fig. \ref{sup_fig_4}. 

\begin{table}[htbp]
\caption{\label{sup_tab_1}
Simulation results of polarized target implosion of SPINSIM and MULTI-IFE.}
\begin{ruledtabular}
\begin{tabular}{lcc}
 &SPINSIM & MULTI-IFE\\
\hline
Minimum fuel radius [$\mu$m] &      56.0      &  66.0 \\
Peak fuel areal density $(\rho R)_{fuel}$ [g/cm$^2$] & 0.0702 & 0.0794\\
Peak fuel center density $\rho_{center}$ [g/cm$^3$] & 4.12  & 10.25 \\
Peak fuel center ion temperature $(T_i)_{center}$ [keV] & 4.67 & 6.61 \\
Peak neutron production rate $R_n$ [10$^{25}$ n/s] &  0.484  & 1.003 \\
Bang time [ns] &  12.85  & 12.76 \\
Burn width [ps] & 240 &  210 \\
Neutron yield Y [10$^{15}$]  &  1.271  &  2.33
\end{tabular}
\end{ruledtabular}
\end{table}

\begin{figure}[htbp]
\includegraphics[width=1\textwidth]{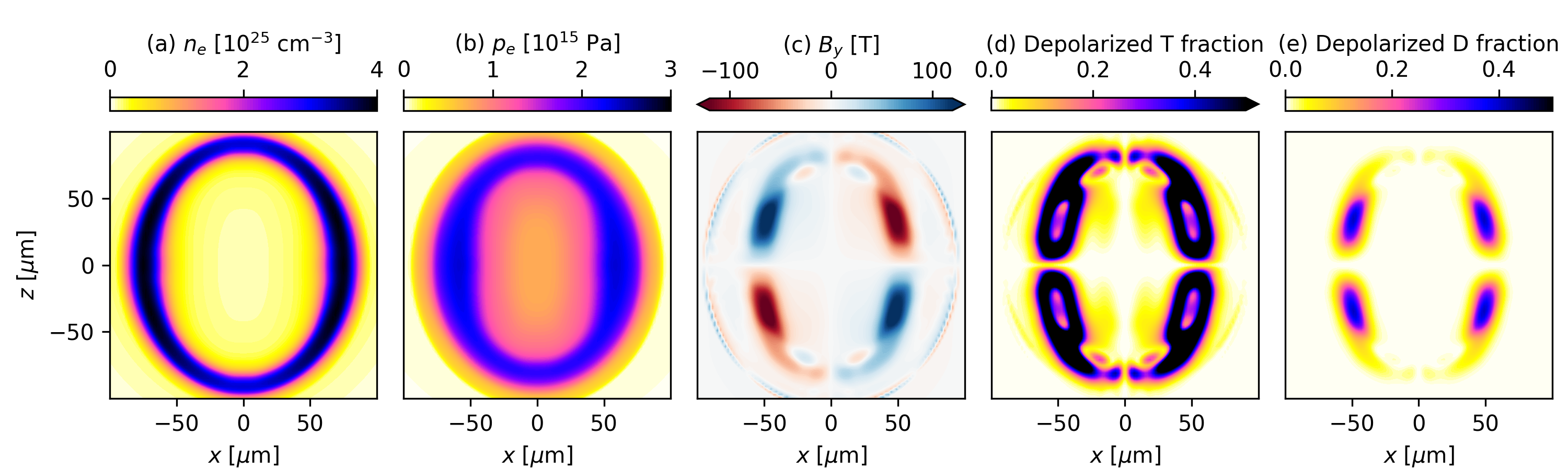}
\centering
\caption{\label{sup_fig_5} SPINSIM simulation results of polarized target implosion with $P_2$ perturbations. Distribution on $y=0$ plane at 12.85 ns of (a) electron density $n_e$, (b) electron pressure $p_e$, (c) out-of-plane magnetic field $B_y$, (d) depolarized triton fraction ($\eta_{11}^T$), (e) depolarized deuteron fraction ($\eta_{11}^D+\eta_{22}^D$). The shell asymmetry amplitude $P_2/P_0$ is 0.21. The initial spins of DT fuel are aligned parallel to the hohlraum axis.}
\end{figure}

The low-mode polar asymmetries can be modeled using Legendre modes \cite{zylstra2015},
\begin{equation}
\label{S35}
R_{shell}(\theta) = R_0\left(1+a_l \sqrt{\frac{2l+1}{4\pi}}P_l(\cos\theta)\right),
\end{equation}
where $R_{shell}$ is the shell radius, $R_0$ is the unperturbed radius, $P_l$ is the Legendre polynomial of order $l$, $a_l$ is the fractional
asymmetry amplitude, which is also referred to as $P_l/P_0$. In SPINSIM, the low-mode polar perturbations are added to the initial radial implosion velocity \cite{chittenden2016} obtained from MULTI-IFE as
\begin{equation}
\label{S36}
v_r(r,\theta) = v_{r0}(r)\left(1+a_l \sqrt{\frac{2l+1}{4\pi}}P_l(\cos\theta)\right),\ \text{for } r > r_{fuel}.
\end{equation}
Here $v_{r0}$ is the unperturbed radial velocity, $r_{fuel}$ is the outer radius of the fuel. Fig. \ref{sup_fig_5} shows the simulation results of polarized target implosion with $P_2$ perturbations. The gradients of electron density and pressure are not in parallel due to the perturbations and magnetic fields are generated through the Biermann battery effect as shown in Fig. \ref{sup_fig_5}(a)-(c). The magnetic fields generated by polar asymmetries are in azimuthal directions with respect to the hohlraum axis. The magnetic fields are strong near the fuel-shell interface because the gradients of electron density and pressure are relatively large. But there are also some small amplitude fields inside the fuel, which can cause some of the tritons close to the hot-spot to be depolarized as shown in Fig. \ref{sup_fig_5}(d). Depolarized deuterons are mostly located near the fuel-shell interface as shown in Fig. \ref{sup_fig_5}(e).

The high-mode perturbations, which are consist of $N$ random Rayleigh-Tayler spikes or bubbles, are also imposed on the implosion velocity
\begin{equation}
\label{S37}
v_r(x,y,z) = v_{r0}(r)\left(1+\sum_m^N a_m\exp\left(-d^2/w_m^2\right)\right),\ \text{for } r > r_{fuel}.
\end{equation}
$a_m\sim\textrm{Uniform}(\left[-0.1,0.1\right])$ is the random perturbation amplitude, $w_m\sim\textrm{Uniform}(\left[0,\frac{\pi}{5}\right])$ is the initial perturbation width, $d$ is the distance from the perturbation center on a unit sphere surface,
\begin{equation}
\label{S38}
d = 2\arcsin\left(\sqrt{\left(\frac{x}{r}-x_m\right)^2+\left(\frac{y}{r}-y_m\right)^2+\left(\frac{z}{r}-z_m\right)^2}/2\right).
\end{equation}
The perturbation center is randomly located on a unit sphere surface, with $z_m\sim \textrm{Uniform}([-1,1])$, $\phi\sim \textrm{Uniform}([0,2\pi))$ and $x_m=\sqrt{1-z_m^2}\cos\phi$, $y_m=\sqrt{1-z_m^2}\sin\phi$.

\begin{figure*}[htbp]
\includegraphics[width=0.75\textwidth]{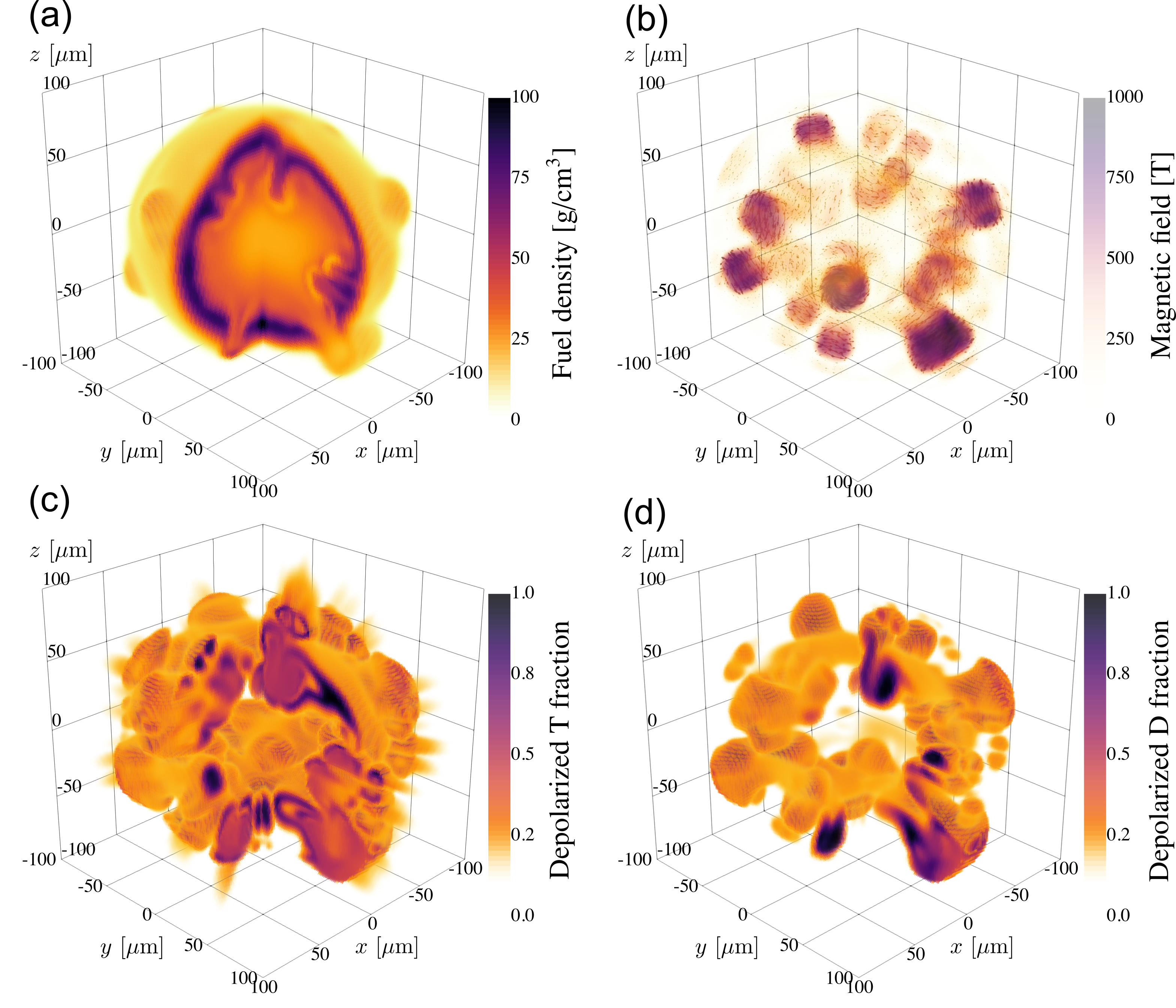}
\centering
\caption{\label{fig1} 3D STHD simulation results of a polarized DT-gas-filled capsule at bang time, (a) fuel density, (b) magnetic field strength, (c) fraction of depolarized tritons ($\eta^T_{11}$), (d) fraction of depolarized deuterons ($\eta^D_{11}+\eta^D_{22}$). The initial fuel polarization is 0.9, the initial spins of DT fuel are perpendicular to the axis of hohlraum. The data in $x,y>0$ region are set to be transparent in (a), (c) and (d) for better visibility}
\end{figure*}

The capsule-only STHD simulation results of a polarized DT-gass-filled capsule are shown in Fig. \ref{fig1}. The capsule is made of a high density carbon (HDC) shell filled with highly polarized DT gas ($p_z^T=p_z^D=p_{zz}^D=0.9$). The outer radius and thickness of the HDC shell are 1040 $\mu$m and 40 $\mu$m respectively. The density of HDC is 3.52 g/cm$^3$ and the density of the DT gas is 4 mg/cm$^3$. The initial temperature of the capsule is 65.65 K. Because the hydrodynamic instabilities and magnetic fields are amplified during the stagnation phase of the implosion \cite{walsh2017}, only the stagnation phase is simulated with SPINSIM. The radiation hydrodynamics code MULTI-IFE \cite{ramis2016} is used to provide the fluid quantities as input data for STHD simulations. The capsule is ablated by radiations with peak temperature of 250 eV and reaches maximum fusion rate after 12.85 ns (bang time). The polar mode-2 ($P_2$) perturbation which forms from low-mode radiation drive asymmetries \cite{zylstra2015,li2021} is added to the implosion velocity of STHD simulation. The asymmetry amplitude of shell radius measured at bang time is $P_2/P_0=0.21$. High-mode perturbations, which rise from the defects of the target, are also added with 64 random RTI spikes and bubbles. The density distribution of the fuel at bang time is shown in Fig. \ref{fig1}(a). The axis of polar asymmetry, which is typically the axis of the hohlraum, is along the $x$ direction. The initial spins of the DT fuel are aligned along the $z$ axis, perpendicular to the axis of hohlraum. The capsule is driven harder around the waist and the shell shape is prolate. RTI spikes and bubbles are developed near the fuel-shell interface. The magnetic fields around the RTI spikes and bubbles are very intense as shown in Fig. \ref{fig1}(b). The spikes can penetrate into the hot-spot and the magnetic fields can cause depolarization of the fuel in the hot-spot. The depolarized tritons and deuterons are mostly distributed near the fuel-shell interface as shown in Figs. \ref{fig1}(c) and (d). More tritons are depolarized than deuterons because the gyromagnetic ratio of deuteron is smaller, which means it takes longer for deuterons to be depolarized by the same magnetic field strength.

\begin{figure}[t]
\includegraphics[width=0.65\textwidth]{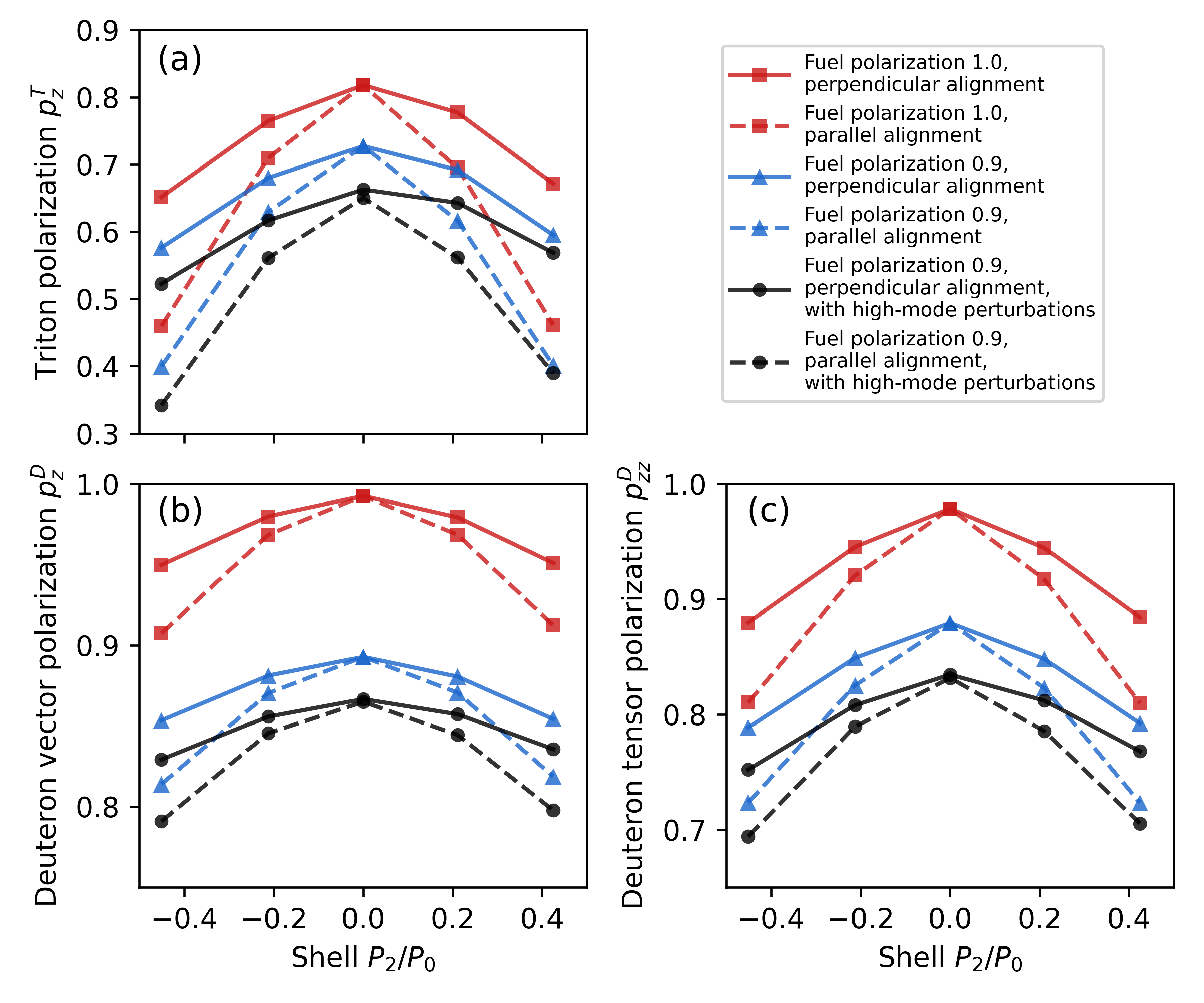}
\centering
\caption{\label{fig2} 3D STHD simulation results of burn rate weighted fuel polarizations at bang time, (a) triton polarization $p_z^T$, (b) deuteron vector polarization $p_z^D$, (c) deuteron tensor polarization $p_{zz}^D$, for different $P_2$ amplitudes with various conditions of initial fuel polarization, spin alignment and high-mode perturbation.}
\end{figure}

The spin alignment of the polarized DT fuels is a key parameter for polarized target design. The magnetic fields generated by polar asymmetries are in azimuthal direction, perpendicular to the axis of hohlraum \cite{walsh2017}. If the initial spins of DT fuel are parallel to the axis of hohlraum, the depolarization is fast because the the spins are perpendicular to magnetic fields. The depolarization of DT ions can be reduced if the initial spins of DT fuel are aligned perpendicular to the axis hohlraum. Fig. \ref{fig2} shows the fuel depolarizations at bang time for different $P_2$ perturbation amplitudes. The fuel polarizations decrease with the increment of the absolute values of $P_2$ amplitudes as shown in Figs. \ref{fig2}(a)-(c). The triton polarizations are smaller than the deuteron polarizations as shown in Figs. \ref{fig2}(a) and \ref{fig2}(b). The perpendicular alignment cases have larger DT polarizations than the parallel alignment cases as shown in Figs. \ref{fig2}(a)-(c).

\begin{figure*}[t]
\includegraphics[width=0.75\textwidth]{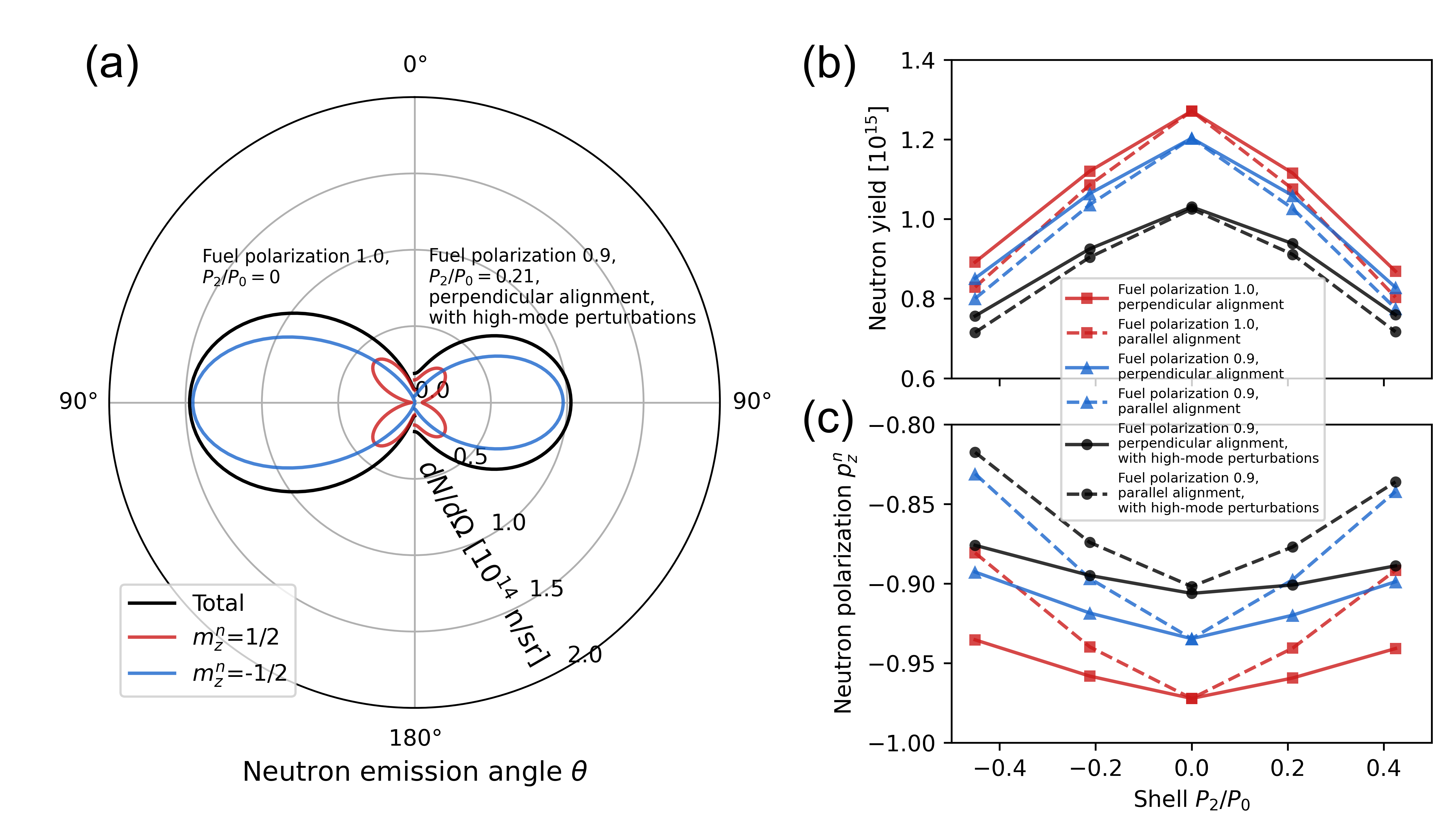}
\centering
\caption{\label{fig3} (a) Neutron angular distributions for polarized capsule implosions with a near-ideal condition (left) and a more realistic condition (right). (b) Neutron yields and (c) neutron polarization $p^n_z$ measured at $\theta=90^\circ$ for different $P_2$ amplitudes with various conditions of initial fuel polarization, spin alignment and high-mode perturbation.}
\end{figure*}

The neutron angular distributions for polarized capsule implosions with a near-ideal condition (with minor asymmetry induced by 3D Cartesian grid and simulation boundaries) and a more realistic condition are shown in Fig. \ref{fig3}(a). The neutron emissions are anisotropic and peaked at $\theta=90^\circ$. The neutrons emitted at $\theta=90^\circ$ are polarized in spin state $m_z^n=-\frac{1}{2}$. The perturbations can reduce the neutron flux and neutron polarization at $\theta=90^\circ$. Neutron yields and neutron polarization $p_z^n$ measured at $\theta=90^\circ$ for different $P_2$ amplitudes with various conditions of initial fuel polarization, spin alignment and high-mode perturbation are shown in Figs. \ref{fig3}(b) and \ref{fig3}(c). The neutron yields and absolute values of neutron polarization decrease with the increasing of absolute values of $P_2$ perturbation amplitudes. Reduction of initial fuel polarization and increment of high-mode perturbations can cause the reduction of neutron yields and absolute values of neutron polarization. Reductions of neutron beam depolarization by $P_2$ perturbations with perpendicular alignment are significant under all conditions as depicted in Fig. \ref{fig3}(c). 

In summary, we have derived the spin transport equation using the density matrix formulation to model the spin transport hydrodynamics of spin-$\frac{1}{2}$ and spin-1 particles. The spin transport equations can be solved in combination with hydrodynamic equations and magnetic induction equations. The solutions of spin transport equations can be used in fusion rate equations to obtain the neutron yield, neutron angular distribution and neutron polarization of polarized DT fusion. The STHD simulation results show that optimized spin alignment of the polarized DT-gas-filled target can reduce the neutron beam depolarization induced by polar mode asymmetries in indirectly driven implosions. The polarized DT-gas-filled target investigated in this Letter is different from the ignition target which contains a high density DT ice layer \cite{betti2016,*hurricane2014,*lepape2018,*zylstra2021, *zylstra2022}. Polarized DT ice are more difficult to produce than polarized DT gas \cite{ciullo2016}. The polarized DT-gas-filled targets can be useful in experiments of nuclear data measurement \cite{zylstra2016,casey2017} and neutron beam production \cite{gatu2018}. It is promising to obtain highly polarized neutron beams using polarized DT-gas-filled target implosions, which will expand the range of applications of fusion based neutron sources.

\acknowledgments{This work was supported by the National
Natural Science Foundation of China (Grant No. 12105193) and Fundamental Research Funds for the Central Universities (Grant No. 2021SCU12119).}
%

\end{document}